\documentclass[12pt,preprint]{aastex}

\shorttitle{Hot Halos of Group Galaxies}

\shortauthors{Jeltema et al.}

\begin{document}

\title{ THE HOT GAS HALOS OF GALAXIES IN GROUPS }

\author{Tesla E. Jeltema\altaffilmark{1,2}, Breanna Binder\altaffilmark{2,3}, and John S. Mulchaey\altaffilmark{2}}

\altaffiltext{1}{Morrison Fellow, UCO/Lick Observatories, 1156 High St., Santa Cruz, CA 95064; tesla@ucolick.org}
\altaffiltext{2}{The Observatories of the Carnegie Institution of Washington, 813 Santa Barbara St., Pasadena, CA 91101}
\altaffiltext{3}{University of California, San Diego, 9500 Gilman Dr., La Jolla, CA 92093}

\begin{abstract}
We use Chandra observations of 13 nearby groups of galaxies to investigate the hot gas content of their member galaxies.  We find that a large fraction of near-IR bright, early-type galaxies in groups have extended X-ray emission, indicating that they retain significant hot gas halos even in these dense environments.  In particular, we detect hot gas halos in $\sim$80\% of $L_K > L_{\ast}$ galaxies.  We do not find a significant difference in the $L_K-L_X$ relation for detected group and cluster early-type galaxies.  However, we detect X-ray emission from a significantly higher fraction of galaxies brighter than $L_{\ast}$ in groups compared to clusters, indicating that a larger fraction of galaxies in clusters experience significant stripping of their hot gas.  In addition, group and cluster galaxies appear to be X-ray faint compared to field galaxies, though a Chandra based field sample is needed to confirm this result.  The near-IR bright late-types galaxies in clusters and groups appear to follow the $L_K-L_X$ relation for early-type galaxies, while near-IR fainter late-type galaxies are significantly more X-ray luminous than this relation likely due to star formation.  Finally, we find individual examples of ongoing gas stripping of group galaxies.  One galaxy shows a 40-50 kpc X-ray tail, and two merging galaxy systems show tidal bridges/tails of X-ray emission.  Therefore, stripping of hot galactic gas through both ram pressure and tidal forces does occur in groups and clusters, but the frequency or efficiency of such events must be moderate enough to allow hot gas halos in a large fraction of bright galaxies to survive even in group and cluster cores.
\end{abstract}


\section{ INTRODUCTION }

X-ray observations of the hot gas halos ($\sim 10^7$ K) surrounding galaxies offer a new window on galaxy evolution and the effects of environment.  In dense environments, like clusters of galaxies, the galaxy populations are known to be quite different from galaxies in the field: they typically have early-type morphologies and little active star formation (e.g. Dressler 1980; Balogh et al. 1997).  Several mechanisms have been proposed to transform galaxies in dense environments which can also act to remove their hot gas halos.  Broadly, these include mergers/tidal interactions with other galaxies (e.g. Spitzer \& Baade 1951; Richstone 1976; White 1978) or the cluster potential (e.g. Merritt 1984; Miller 1986; Byrd \& Valtonen 1990), ram-pressure or viscous stripping through interaction with the intracluster medium (ICM; Gunn \& Gott 1972; Nulsen 1982), or evaporation by the hot ICM (Cowie \& Songaila 1977).  Typically, ram-pressure stripping is thought to be more effective in high density cluster cores (e.g. Gunn \& Gott 1972), while galaxy-galaxy mergers are thought to be more common in less massive galaxy groups where the relative galaxy velocities are lower (e.g. Barnes 1985; Aarseth \& Fall 1980; Merritt 1984).  However, recent simulations and observations show that ram pressure can be effective at stripping the hot gas, though not the cold gas, from some galaxies even in low-mass groups (Kawata \& Mulchaey 2007; Rasmussen, Ponman, \& Mulchaey 2006).

In spiral galaxies, the hot gas halo provides a reservoir of gas which can cool and replenish the cold gas supply to fuel star formation.  If this gas is stripped from galaxies in clusters and groups, star formation will be quenched after a few gigayears when the galaxy exhausts its gas supply.  Referred to as ``strangulation'' or ``starvation'', this mechanism has been proposed to explain the observed lack of star-forming galaxies and the transformation of spirals to S0 galaxies (as the disk fades) in clusters (e.g. Larson et al. 1980; Bekki et al. 2002).  In elliptical galaxies, hot gas halos are formed and enriched through stellar mass lost from evolved stars and possibly through gas accreted from the local environment (e.g. Mathews 1990; Mathews \& Brighenti 2003).  Observations with \textit{Einstein} showed that hot gas halos surrounding ellipticals are common (e.g. Forman, Jones, \& Tucker 1985; Kim, Fabbiano, \& Trinchieri 1992).  A correlation was found between the X-ray luminosity of early-type galaxies and optical or near-IR luminosity, but with surprisingly large scatter (e.g. Forman et al. 1985; Trinchieri \& Fabbiano 1985; Canizares, Fabbiano, \& Trinchieri 1987; O'Sullivan, Forbes, \& Ponman 2001; Ellis \& O'Sullivan 2006).  In addition to several intrinsic galaxy properties (dark matter profile, rotation, supernovae rate), environment and ram-pressure stripping in dense environments are among the proposed factors to explain this scatter (see review by Mathews \& Brighenti 2003).  The stripping of gas from cluster and group galaxies may also be an important contribution to the enrichment of the ICM (Byrd \& Valtonen 1990; White 1991; Schindler et al. 2005; Domainko et al.2006; Kapferer et al.2007).

How effective are stripping mechanisms at removing the hot gas from typical galaxies in clusters?  In lower density groups?  
Based on the Millenium Simulations, Br\"{u}ggen \& De Lucia (2007) predict that nearly all galaxies in the cores of clusters experienced strong ram pressure in their histories.  In addition, observational evidence for hot gas stripping has been found for $\sim 10$ galaxies in clusters or groups in the form of long X-ray tails trailing the galaxy (Forman et al. 1979; Rangarajan et al. 1995; Irwin \& Sarazin 1996; Biller et al. 2004; Trinchieri et al. 1997; Kim et al. 2007; Sakelliou et al. 2005; Wang et al. 2004; Scharf et al. 2005; Machacek et al. 2005a, 2005a, 2006, 2007; Rasmussen et al. 2006; Sun \& Vikhlinin 2005; Sun et al. 2006).  However, a recent \textit{Chandra} archival survey of galaxies in 25 clusters by Sun et al. (2007; S07) found that most bright cluster galaxies ($>$ 60\% of galaxies above 2$L_{\ast}$) host hot halos.  Supporting these observations, recent hydrodynamic simulations by McCarthy et al. 2007 show that for typical galaxy structures and orbits, galaxies falling in to a massive group/poor cluster maintain $\sim$30\% of their hot halo gas after 10 Gyr, and the amount of stripping depends signficantly on the orbital history.

The superb resolution of \textit{Chandra} is invaluable in the study of galaxy halos, allowing a separation between extended X-ray halos and point source AGN.  Previous studies which did not separate these two components have given mixed results regarding the relationship of the X-ray emission from galaxies to environment.  For example, some \textit{ROSAT} studies did not find a significant trend in the $L_K-L_X$ or $L_B-L_X$ relations with environment (O'Sullivan et al. 2001; Helsdon et al. 2001; Ellis \& O'Sullivan 2006), while other authors found that galaxies in regions of high local galaxy density have lower $L_X/L_B$ (White \& Sarazin 1991; Henriksen \& Cousineau 1999), and Brown \& Bregman (2000), whose sample included several brightest group galaxies (BGG), found the opposite trend for galaxies in higher density environments to have higher $L_X/L_B$.  Recent \textit{XMM-Newton} and \textit{Chandra} observations of the X-ray luminosity function of Coma cluster galaxies indicate that these are X-ray underluminous compared to field galaxies (Finoguenov et al. 2004; Hornschemeier et al. 2006).

In this paper, we extend the \textit{Chandra} study of hot galaxy halos to lower density environments through a systematic study of archival observations of 13 galaxy groups in order to re-examine the effects of environment on hot gas halos.  
Our sample is presented in \S2, the data reduction in \S3, and our results in \S4. These include the $L_K-L_X$ relation and halo temperatures of early-type galaxies (\S4.1), the detection rate (\S4.2), detections of late-type galaxies (\S4.3), and examples of X-ray tails and mergers (\S4.4).  Throughout the paper, we assume $H_0=70$ km s$^{-1}$ Mpc$^{-1}$.

\section{ SAMPLE }

\subsection{Group Catalog}

Groups were selected from the low-redshift group catalog in Mulchaey et al. (2003).  Groups were required to have deep archival \textit{Chandra} observations of at least 30 ksec before flare filtering.  We also required that the \textit{Chandra} field of view cover at least a radius of 60 kpc to ensure that we are probing beyond the central galaxy, which led to lower limits on the group redshift of $z=0.012$ for ACIS-S observations and $z=0.006$ for ACIS-I observations.  These two criteria gave us a list of 12 groups; to this list we add HCG 16 which has a shallower exposure, but a low redshift and is known to host galaxies with extended X-ray emission.  Our group sample includes the full range of group types, including compact groups, X-ray faint, and X-ray luminous groups.  Table 1 lists the \textit{Chandra} observations used, the flare free exposure times, the group redshifts, velocity dispersions, and X-ray luminosities.  Group luminosities are taken from Mulchaey et al. (2003).  We choose to list ROSAT luminosities, because the larger field if view allows the group luminosity to be probed outside the core.  
Group velocity dispersions are taken from Osmond \& Ponman (2004) where possible or calculated from the catalog of group members listed in the NASA Extragalactic Database (NED).

\subsection{ Group Galaxies }

For each group, we created a list of member galaxies using NED.  We include all galaxies with velocity offsets from the central galaxy within three times the group velocity dispersion which also fell within the \textit{Chandra} field.  Typically, the BGG lies near the center of the group potential (Zabludoff \& Mulchaey 1998), and we found that changes in the adopted central velocity did not affect the galaxy catalogs.  For ACIS-I observations, we limited the galaxy catalog to galaxies falling on one of the four I chips, while for ACIS-S observations, we included only galaxies falling on the S2, S3, and S4 chips.  In this way, we consider only detector regions where the PSF is relatively small.  Initially, we consider all galaxies with redshifts placing them within the group when matching detected X-ray sources; however, in the fits that follow we limit the catalog to galaxies with $K_s$ band luminosities of $L_{Ks} > 10^{10.45} L_{K,\odot}$ ($K_{\odot} = 3.39$).  This luminosity limit ensures that all galaxies would have apparent magnitudes above the 2MASS limit of $K_s = 13.5$ mag across the redshift range covered by our sample.
Where possible $K_s$ band luminosities are taken from 2MASS and galaxy morphologies from NED.  For two groups, N5171 and N6269, additional galaxy morphologies not available in NED were classified based on SDSS images.  For a few galaxies lying in close pairs, 2MASS magnitudes were not available, and these were determined by converting $B$ or $r$ band magnitudes to $K_s$ band using the spectral energy distributions of Coleman, Wu, \& Weedman (1980).  $K_s$ band luminosities are extinction corrected using the NED values (Schlegel, Finkbeiner, \& Davis 1998).  We excluded from the list galaxies lying at the center of groups with diffuse X-ray emission, as these lie at the center of the group potential and may have X-ray properties significantly different from satellite galaxies.  The brightest group galaxies were included for HCG16, NGC3557, NGC5171, and HCG90.  NGC5171 contains diffuse, group scale X-ray emission, but this emission is not centered on any of the bright galaxies in this system (Osmond, Ponman, \& Finoguenov 2004).  The other three groups do not have obvious group scale X-ray emission in the \textit{Chandra} observations.

\section{ DATA REDUCTION }

The data were prepared using CIAO 3.2.1 and CALDB version 3.0.3 following the standard data processing.  We chose to reprocess the data starting from the level 1 file, including redetecting hot pixels and afterglow events using the latest tools, applying the newest gain file, and destreaking the S4 chip.  For observations at a focal plane temperature of $-120^{\circ}$C, we also applied the CTI and time-dependent gain corrections.  The ACIS-S observation of HCG62 was taken at a focal plane temperature of $-110^{\circ}$C, for which no CTI correction was available; however, for this detector and the low-resolution spectroscopy in this work, we expect CTI to have little effect.  We kept only events with \textit{ASCA} grades of 0, 2, 3, 4, and 6 and a status of zero\footnote{\textit{Chandra} Proposers' Observatory Guide http://cxc.harvard.edu/proposer/POG/, section ``ACIS''}, and in the case observations performed in VFAINT (VF) mode, the additional background cleaning for VF mode data was also applied.  We removed background flares from the data following the prescriptions of Markevitch et al. (2003).
The filtering excluded time periods when the count rate, excluding point sources and group emission, was not within 20\% of the quiescent rate.  For ACIS-S observations, we detected flares in the 2.5-7 keV band, while for ACIS-I the 0.3-12 keV band was used.  Where possible for ACIS-S observations, we detected flares on the other back-illuminated CCD, S1 in order to have a significant area free of group emission.

Sources were detected in the 0.5-2 keV band using wavdetect, a wavelet source detection tool in CIAO, using wavelet scales of 1, 2, 4, 8, and 16 pixels.  Here the detection threshold was set to give approximately one false detection per image.  Detected X-ray sources are matched against our catalogs of group galaxies; matches within 2'' are listed in Table 2.  We also compared the X-ray source positions to the Digital Sky Survey (DSS) images, and all of these sources lie near the center of their host galaxies.  To distinguish between X-ray emission from a hot gas halo and from an AGN, we examine source extent in the 0.5-2 keV image.  X-ray sources were fit with a Gaussian profile and compared to the PSF at the galaxy location.  Sources are considered to be extended if their FWHM is more than 20\% larger than the FWHM of the PSF.  We also examined extent by subtracting the PSF normalized to the central surface brightness of the source from the X-ray profile, and we found that all sources classified as extended were at least three sigma above background after the PSF subtraction.  Sources detected as extended are labeled 'E' in Table 2, while point-like sources are labeled 'P'.  For sources with less than 40 counts we found that we could not reliably determine source extent, and these sources are labeled 'U' for unknown.  Undetected galaxies above our $K_s$ band luminosity cut of $L_{Ks} > 10^{10.45} L_{K,\odot}$ are listed in Table 3.  We do not attempt to stack galaxies, because the low galaxy densities in groups mean that we do not typically have many galaxies per group.

\subsection{Spectroscopy}

For detected sources, we extracted source spectra in circular regions extending to where the surface brightness profile reaches the background level.  Local background spectra were extracted from typically annular regions immediately surrounding the source while avoiding any nearby point sources.  A local background of this type will include any additional background from the diffuse group emission at the location of the source.  Response files (RMFs and ARFs) were determined for the location of the source, and the source spectra were binned to have 15 counts per bin.

Even for extended sources where we believe that a hot gaseous halo is present, there may by some contribution to the X-ray emission from an AGN or low mass X-ray binaries (LMXB) in the galaxy.  In a couple of the brighter galaxies, we, in fact, detect both an extended component and a central point source. Therefore, we fit sources to a two component model: a thermal mekal model plus a power law.  Spectra were fit in the 0.5-7 keV range to better constrain the hard component.  Here we follow a methodology similar to that used by Sun et al. (2007), in order to directly compare our results.  In the spectral fitting, the column density was fixed at galactic and the metallicity was fixed at 0.8 solar (Anders \& Grevessa 1989).  In general, for extended sources detected at $>10 \sigma$ and with more than 150 counts (in the detection band of 0.5-2 keV), we fit both the temperature and photon index.  For sources with more than 60 counts (typically detected at $>5 \sigma$), we fixed the photon index at 1.7 and fit only the temperature and the two normalizations.  If the parameters were found to be unconstrained at the $3 \sigma$ level, we fixed an additional parameter.  In Table 2, we report the unabsorbed luminosity of the thermal component, both in the 0.5-2 keV band and bolometric.  The bolometric luminosities can vary significantly with the choice of spectral parameters, but the 0.5-2 keV band are significantly less sensitive.
For spectra with three or more free parameters, the luminosity errors were determined using Monte Carlo Markov chains and the luminosity reported is the median luminosity of the chain; otherwise, errors were determined from the one sigma errors on the free parameters.

For sources detected with less than 60 counts we found it difficult to independently constrain the normalizations of the thermal and power law components.  Two extended sources in this category (NGC0382 and ARK066) have very soft hardness ratios (H(2-10 keV)/S(0.5-2 keV) $<$ 0.05) inconsistent with having a hard component.  For these sources, we assumed that all of the emission was thermal and fit a thermal only model.  One detected point source, the Seyfert 2 galaxy NGC0833, was found to have a very hard hardness ratio (H(2-10 keV)/S(0.5-2 keV) $>$ 1.0) and is discussed below.  For all other detected sources with less than 60 counts (point or extended), we place an upper limit on the luminosity from a gaseous halo using the one sigma upper limit from a thermal only model with a fixed temperature of 0.7 keV.

Three sources were consistent with having emission solely from an AGN.  These are two known Seyfert 2 galaxies, NGC0833 and NGC7172, as well as a bright point source in NGC0742.  NGC7172 also appears to have an extended component, but the AGN is very bright and we were unable to get a constrained fit to the thermal component even when limiting the energy range to 0.5-2 keV and removing the central region from the fit.  For these sources we determined an upper limit on any thermal emission by first fitting the source spectrum to a power law model (including fitting the absorption for the two Seyfert galaxies) and then adding a thermal component with fixed kT $=0.7$ keV.  The upper limit on the luminosity is set to the three sigma upper limit from this thermal component.  For undetected sources, we follow the same procedure as S07.  We extract spectra in a 3 kpc region at the position of the galaxy and place an upper limit on the luminosity using a fixed thermal model with kT $=0.7$ keV and a normalization given by the Poisson $3 \sigma$ upper limit on the count rate in this region.

Tables 2 and 3 list the redshifts, $K_s$ band and X-ray luminosities, optical morphologies, and radii of detected and undetected galaxies.  For detected galaxies, the temperature and and spectral index are also listed.  Radius refers to the distance of the galaxy from the peak of the group X-ray emission for those groups with significant diffuse emission or the distance from the central galaxy in X-ray faint groups (HCG16, NGC3557, and HCG90).

\section{ RESULTS }

\subsection{ Early-Type Galaxies: Scaling Relations }

The X-ray luminosity of galaxies is known to correlate with optical/near-IR luminosity, although with large scatter, indicating a correlation between X-ray emission and galaxy stellar mass.  Here we consider the correlation between $L_X$ and $L_K$, because $K_s$ band luminosity is more tightly correlated with stellar mass than bluer bands.  In Figure 1 we show the $L_K-L_{0.5-2 keV}$ relation for detected early-type galaxies in our groups compared to early-type galaxies in clusters from S07.  For consistency with S07, we define early-type as galaxies classified as Sa or earlier.  We have removed galaxies at the center of clusters (with $r=0$ in Table 2 of Sun et al. 2007) from the S07 sample, and here we only consider galaxies identified by their criteria 2 (i.e. spatially extended).  S07 also identify halos based on the presence of an iron L-shell hump in the spectrum (criterion 1) or a photon index indicating a soft spectrum (criterion 3).  With the exception of two AGN, the non-extended sources in our sample have fewer than 60 counts preventing detailed spectral analysis of these sources; their hardness ratios also do not indicate that they are particularly soft.  For these reasons, using additional spectral criteria to identify galaxy halos would not significantly change our sample.

We fit the $L_K-L_X$ relations for the group and cluster samples as log$(L_{0.5-2 keV}) = A + B$log$(L_K/10^{11}L_{K,\odot})$ using the bisector modification to the BCES method in Akritas \& Bershady (1996).  The results are shown in Figure 1 and Table 4; errors on these fits are determined using 10,000 bootstrap trials.  The best-fit to the group relation is steeper than the cluster relation, but the same within the errors.  For groups, we find $B = 2.74 \pm 0.63$, and for clusters, $B = 1.57 \pm 0.28$.  The normalizations of the two relations are similar.  We, therefore, find no indication that galaxies in clusters are deficient in hot gas compared to galaxies in groups.  If we include in the cluster sample all galaxies identified by S07 as containing halos, we find very similar results.  Applying a survival analysis using the Buckley-James algorithm to include the upper limits in both samples (Feigelson \& Nelson 1985) also does not significantly change the fits.  All fits are listed in Table 4.

\begin{figure}
\epsscale{0.8}
\plotone{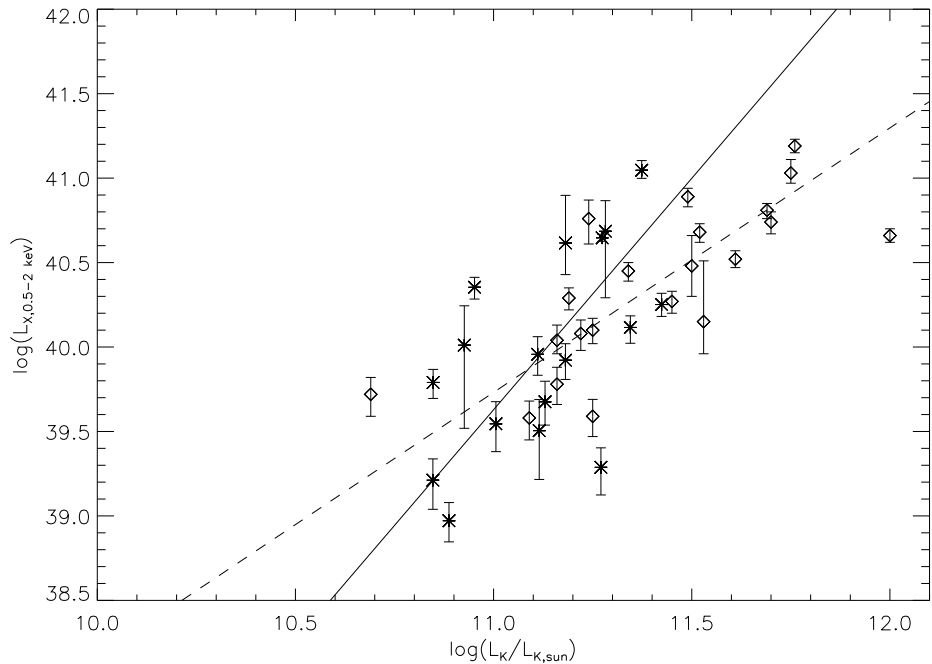}
\caption{ $L_K-L_{0.5-2 keV}$ relation for early-type galaxies with extended X-ray halos in groups (asterisks) and the cluster sample of S07 (diamonds).  X-ray luminosities include only the thermal component.  Also shown are the best fit relations to these samples (solid line: groups, dashed line: clusters).  The best-fit group relation is steeper, but statistically consistent with the cluster relation. }
\end{figure}

We also tried subdividing our detected galaxy sample based on radius from the group center normalized to $r_{500}$ or total group X-ray luminosity, but with our small sample size, we did not find any significant differences in the $L_K-L_X$ relation.  We also did not find a significant radial trend when adding the cluster galaxies from the S07 sample.  However, a radial trend might be difficult to detect, because the \textit{Chandra} observations only cover the cores of these systems.

Finally, we compare our results to the X-ray properties of early-type galaxies in the field from the \textit{ROSAT} study of Ellis \& O'Sullivan (2006).  Here we consider only galaxies classified by Ellis \& O'Sullivan (2006) as being in the field environment.  \textit{ROSAT}, unfortunately, did not have the resolution to determine whether galaxies at these distances were extended or point-like, but currently no large sample of field galaxies studied with \textit{Chandra} exists.  Samples of nearby early-type galaxies observed with \textit{Chandra} include almost exclusively members of clusters or groups with $\sim6$ field galaxies in our redshift range (e.g. Athey 2007).  Ellis \& O'Sullivan (2006) assume a fixed thermal model with kT $=1.0$ keV and solar abundance.  We use a constant conversion factor of 0.52 (from PIMMS) to translate their bolometric luminosities to a thermal model with kT $=0.7$ keV and metallicity of 0.8 solar in the 0.5-2 keV band to better match our results.  We convert their bolometric luminosities to 0.5-2 keV band luminosities, because these are much less sensitive to the assumed spectral parameters.  The results are plotted in Figure 2.  

We find the $L_K-L_{0.5-2 keV}$ relation for field galaxies has a similar slope to the group and cluster relations (Table 4), but a significantly higher normalization.  However, the Ellis \& O'Sullivan (2006) luminosities assume that all of the X-ray emission is thermal gas.  Known AGN are excluded from their sample, but their luminosities will include the contributions from LMXBs and low-luminosity AGN.  To test the effect that this has on the $L_K-L_{0.5-2 keV}$ relation, we recalculate the luminosities of all of our detected group galaxies, including those detected as point sources but removing galaxies identified in NED as Seyferts or LINERs, assuming all of the emission is thermal with kT $=0.7$ keV and a metallicity of 0.8 solar and fitting only the spectral normalization.  The resulting group $L_K-L_{0.5-2 keV}$ relation (Figure 2 and Table 4) is flatter and has a slightly higher normalization, but is still significantly below the field relation.  It appears that group galaxies contain significantly less hot gas than field galaxies, indicating that some stripping has occurred in these dense environments.  However, this result should be checked with a \textit{Chandra} study of field galaxies as other differences like contamination from background point sources may effect the \textit{ROSAT} luminosities.  

We note here that for the three galaxies (NGC3557, NGC7173, and NGC7176) that overlap between our sample and that of Ellis \& O'Sullivan (2006), we find the ratios of their luminosities to ours to be 0.48, 40, and 17.  If we instead assume that all of the emission is due to a thermal component for the Chandra spectra as above, these ratios become 0.98 (NGC3557), 32 (NGC7173), and 9.4 (NGC7176).  For NGC3557, the ROSAT and Chandra luminosities are very consistent.  NGC7173 and NGC7176, however, presents a more complicated case as they are part of a three galaxy merger/interaction in the group HCG90 (see \S 4.4).  Using \textit{Chandra's} resolution we are able to separate these galaxies, and we conservatively take source regions just around the individual galaxies, treating the surrounding stripped/common halo gas as background.  In the ROSAT image, these galaxies are not clearly separated, particularly the merging pair NGC7176/NGC7174, and the background was taken from well outside this region.  If we consider the total X-ray emission in the region around these three galaxies as a single halo, we find a \textit{Chandra} luminosity within 40\% of the Ellis \& O'Sullivan (2006) luminosity, making them consistent within the luminosity errors, and we suspect that the higher \textit{ROSAT} luminosities are due to contamination from the neighboring galaxies and gas.  S07 similarly find a discrepancy between their luminosities and those of Ellis \& O'Sullivan (2006) for a few cluster galaxies which have significant contamination from the ICM in the \textit{ROSAT} observations.  This type of contamination should not be present for the field sample and should not bias our results.

\begin{figure}
\epsscale{0.8}
\plotone{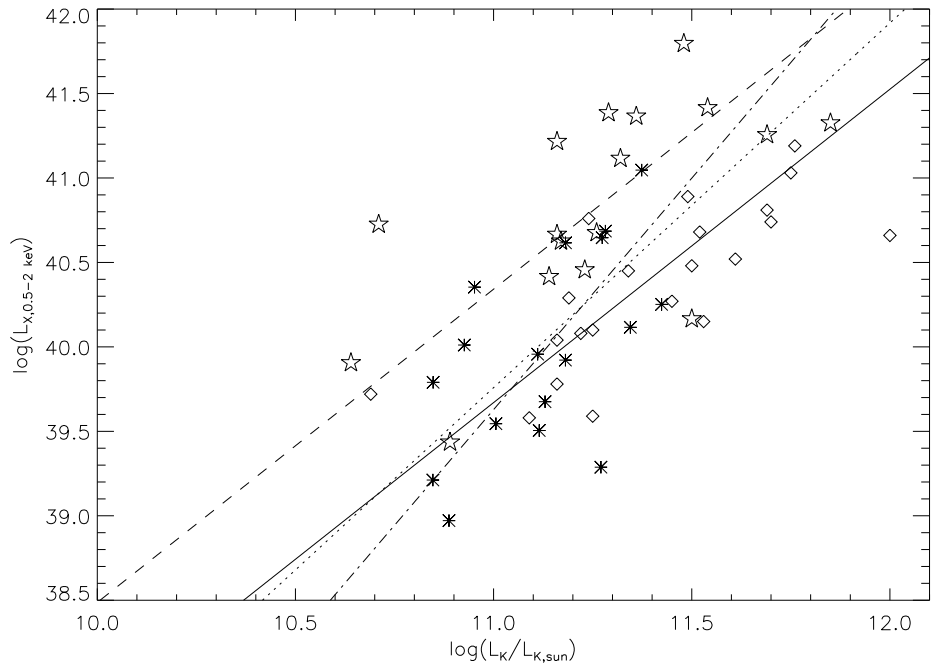}
\caption{ $L_K-L_{0.5-2 keV}$ relation for early-type galaxies in groups (asterisks) and clusters (diamonds; S07) compared to early-types in the field (stars; Ellis \& O'Sullivan 2006).  The cluster and group samples include only galaxies detected as extended.  The best-fit relations are shown for group galaxies only (dot-dashed line), group and cluster galaxies (solid line), and field galaxies (dashed line). We also show the best-fit relation for group galaxies if we assume a thermal only model as was assumed for the field sample (dotted line). }
\end{figure}

For those galaxies where we were able to fit temperatures, we also compare halo temperature to near-IR luminosity (Figure 3a).  Similar to S07, we find no significant correlation between temperature and $L_K$.  However, group galaxies are significantly cooler than cluster galaxies at a given $L_K$.  We find the same trend when comparing to stellar velocity dispersion (Figure 3b).  Stellar velocity dispersions are taken from HyperLeda (Prugniel \& Simien 1996) where possible with three additional values added from Wegner et al.(1999) and Wegner et al.(2003).  In Figure 3b, the group galaxies tend to have lower temperatures at the same stellar velocity dispersion compared to cluster galaxies, which leads to typically higher values of $\beta_{spec}$ ($\mu m_p \sigma^2_{\ast}/kT$).  $\beta_{spec}$ gives the ratio of the energy in stars to gas.  S07 find that cluster galaxies typically have $\beta_{spec} < 1$, implying an overheating of the gas relative to the stars.  The low velocity dispersion group galaxies in our sample also tend to have low $\beta_{spec}$, but above $\sigma_{\ast} \sim 200$ km s$^{-1}$, the group galaxies have $\beta_{spec} \ge 1$.  More massive group galaxies, therefore, appear to have undergone less heating relative to their stars.  David et al. (2006) find similarly low halo temperatures for a sample of low-luminosity early-type galaxies, many of which are in groups, but they find typical $\beta_{spec}$ values less than unity.

\begin{figure}
\epsscale{0.8}
\plotone{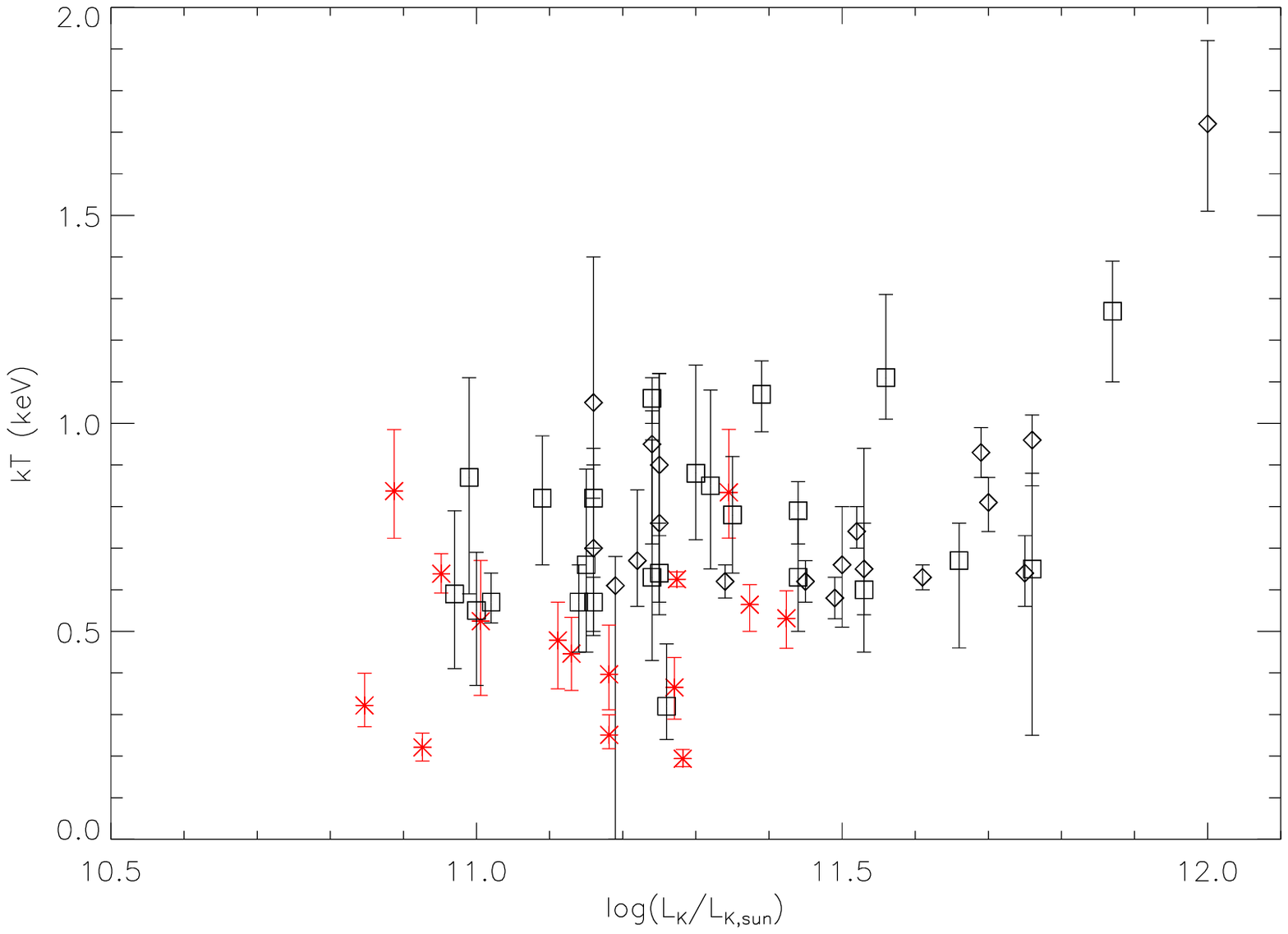}
\plotone{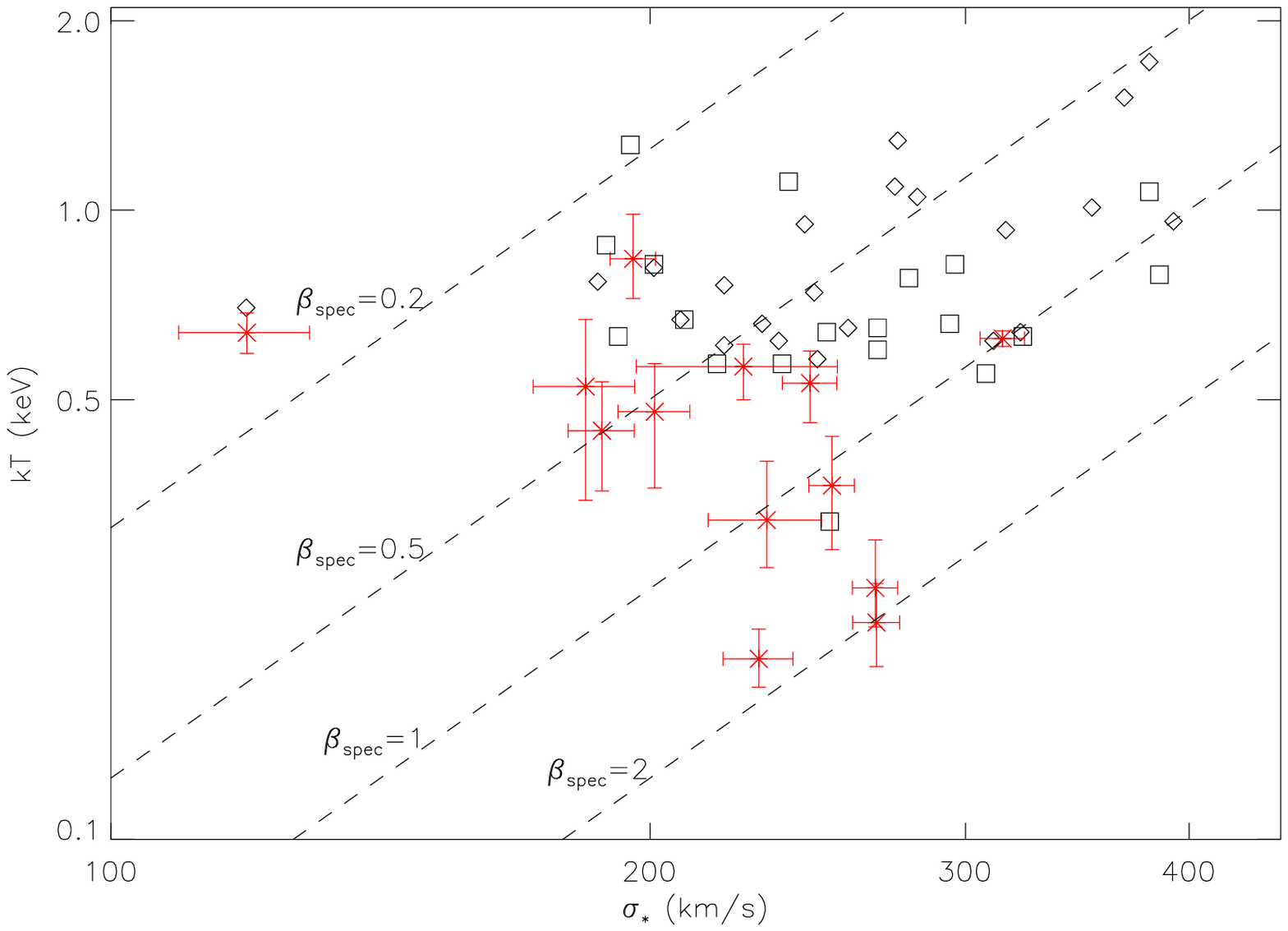}
\caption{ Top: $L_K$ versus halo temperature for early-type galaxies in groups (red asterisks, this paper) and clusters (diamonds: S07 extended, squares: S07 non-extended).  Bottom: Halo temperature versus stellar velocity dispersion.  Also shown are dashed lines indicating the range of $\beta_{spec}$. }
\end{figure}

\subsection{ Detection Rate }

It is clear from the number of detections that many galaxies in clusters and groups maintain at least some of their hot gas halos, but what fraction of bright galaxies in these environments do these represent?  Here we will consider only galaxies with $K_s$-band luminosities greater than $L_{\ast}$.  We take $L_{K\ast} = 10^{11.08} L_{K,\odot}$ following S07.  Above $L_{\ast}$, we detect 11 of 14 galaxies in our groups or 79\%.  To compare to the detection fraction in clusters, we conservatively take all individual detections in the S07 sample, even if they were not found to be extended.  The S07 sample also contains halos detected through the stacking of 2-4 galaxies.  In section 2.1, we included these stacked sources as detections, but here we take them as non-detections for consistency with our analysis.  We find a cluster halo detection fraction above $L_{\ast}$ of $39/90 = 43\%$.  Using a bootstrap resampling of our group galaxies to test the significance of this difference, we find that only 0.4\% of trials give detection fractions as low as that found in clusters.

However, galaxy detection depends on the depth of the observation as well as the background (including cluster/group emission) at the location of the galaxy.  To examine more closely the nature of non-detections, in Figure 4 we plot the $L_K-L_X$ relation of both detections and upper limits in the group (Fig. 4a) and cluster (Fig. 4b) samples compared to a joint fit to the detected extended sources in both samples.  Also shown are the 3$\sigma$ limits on the normalization of the fit and a solid vertical line marking $L_{\ast}$.  It can be seen that above $L_{\ast}$ most of the non-detections have upper limits on their X-ray luminosities which are not significantly above the fit, indicating that the X-ray observations are typically deep enough to detect these galaxies.  If we conservatively consider only non-detections whose upper limits fall below the 3$\sigma$ limits on the fit, we find 1 significant non-detection out of 14 galaxies more luminous than $L_{\ast}$ in groups and 16 out of 90 significant non-detections in clusters.  Again, significantly fewer galaxies are detected in clusters (probability of 7.4\% from bootstrap trials).  While we find that the relationship between X-ray luminosity and $K_s$-band luminosity for detected galaxies in groups and clusters is similar, a smaller fraction of bright galaxies is detected in clusters.  This result indicates that while some X-ray halos survive in clusters a higher fraction of galaxies in the dense cluster environment have experienced significant gas stripping versus galaxies in groups.

\begin{figure}
\epsscale{0.8}
\plotone{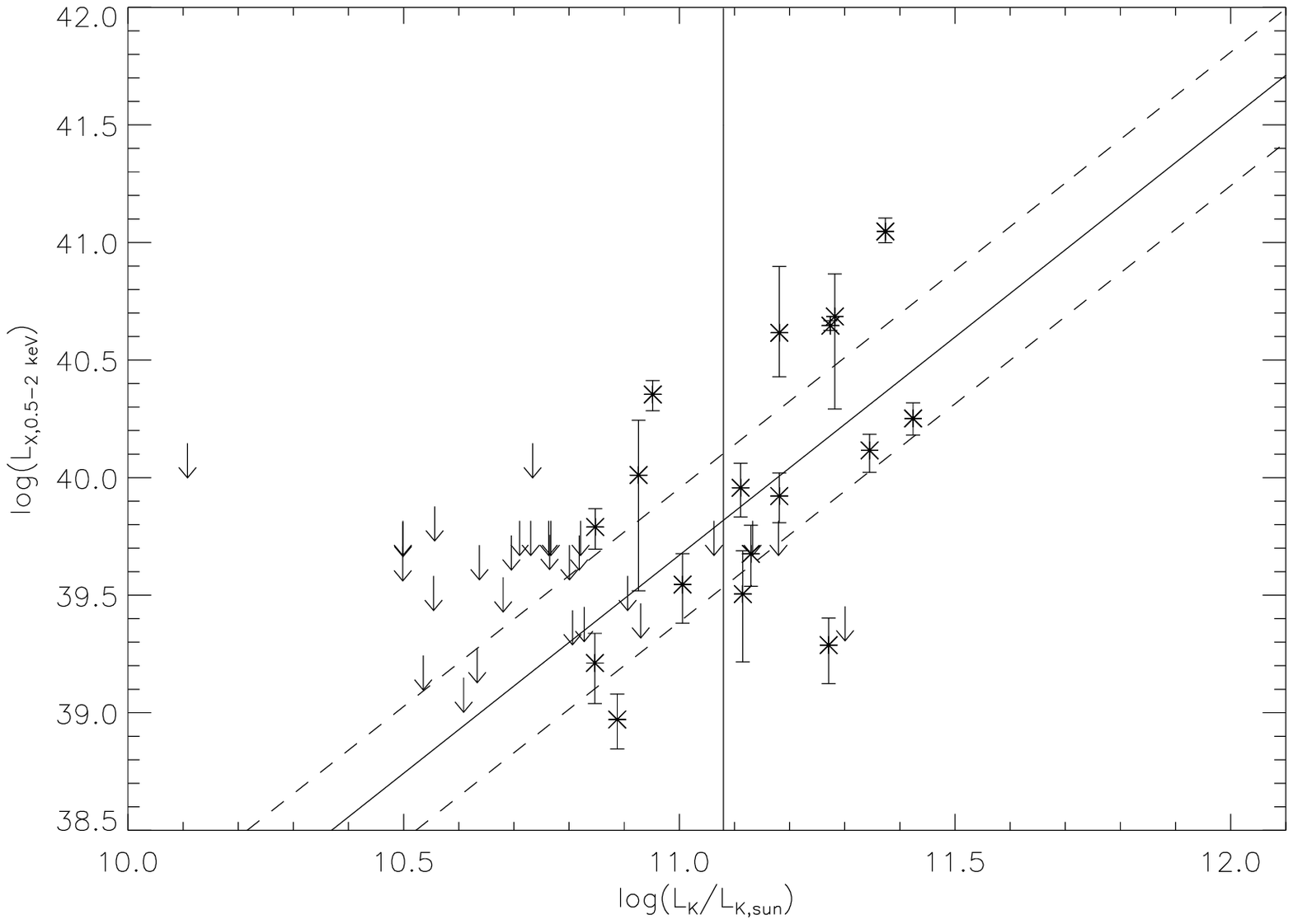}
\plotone{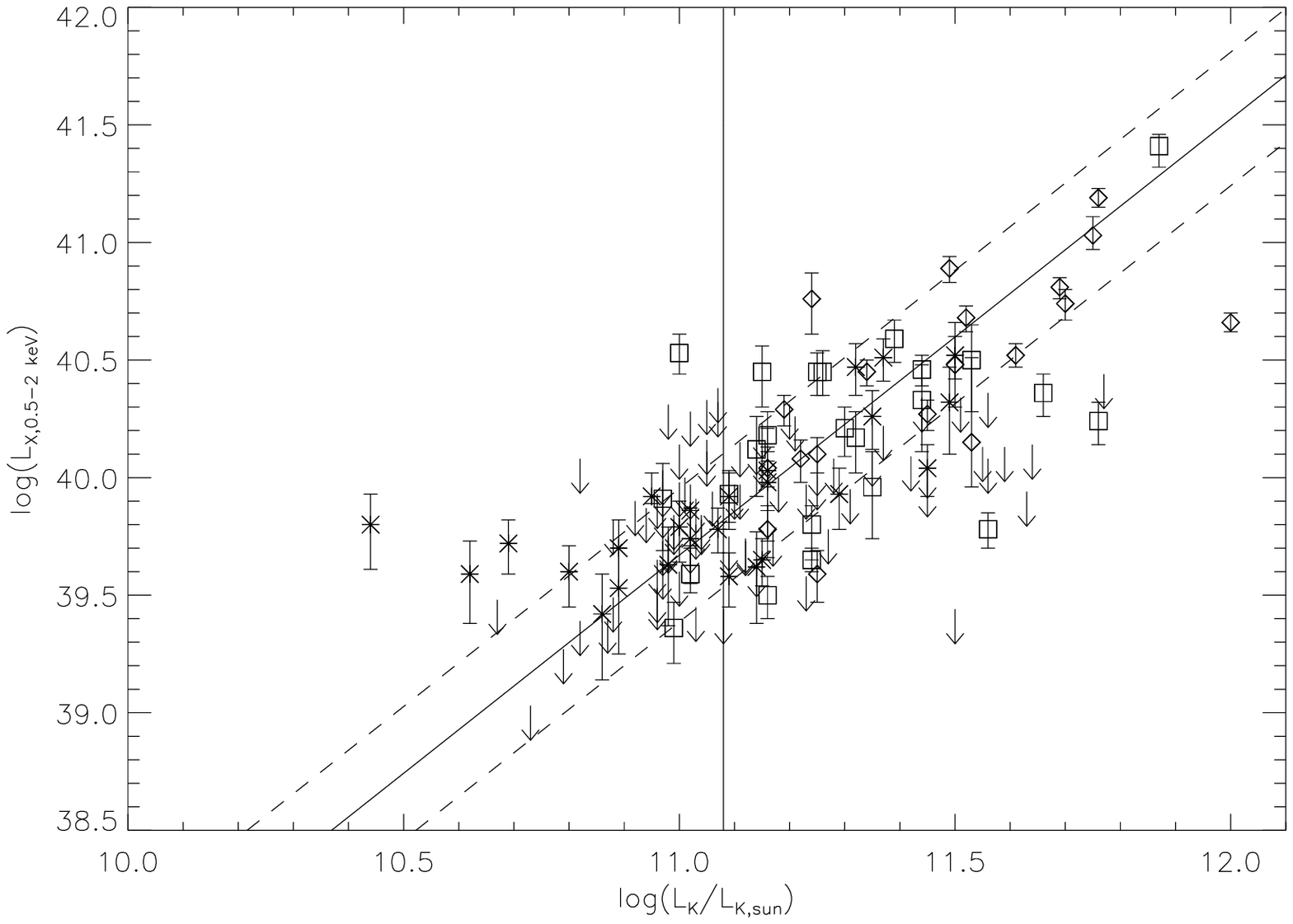}
\caption{ Top: $L_K-L_{0.5-2 keV}$ relation for early-type galaxies in our group sample (asterisks) including upper limits for undetected sources.  Bottom: $L_K-L_{0.5-2 keV}$ relation for early-type galaxies in the S07 cluster sample including upper limits.  Diamonds indicate detected, extended sources, squares indicate halos identified by S07 based on their spectra, and asterisks indicate sources detected by S07 through stacking.  In both plots, the best-fit relation to the group plus cluster extended galaxy sample is shown along with the 3$\sigma$ errors on the normalization, and a solid vertical line marks $L_{K\ast}$. }
\end{figure}

In addition, there are a few detected galaxies in the group and cluster samples whose luminosities place them significantly below the $L_K-L_X$ relation.  For example, we find 1 and 3 galaxies with X-ray luminosities a factor of five or more below what would be expected from the $L_K-L_X$ relation in the group and cluster samples, respectively.  We note that the very underluminous detection in our sample is NGC7176, a galaxy that forms part of a merger between three galaxies in the center of the X-ray faint group HCG90.  This system shows signs of interaction in both the X-ray and optical images (see \S4.4).  The significant non-detection in our sample lies at a fairly large radius of 289 kpc from the center of the X-ray bright group NGC4073.

\subsection{ Late-Type Galaxies }

Our sample also includes 11 late-type galaxies, 4 of which are detected as extended in X-rays.  S07 similarly detect extended X-ray sources for 7 of 22 late-type galaxies in their sample with an additional two detected as halos through their spectra.  We plot the $L_K-L_X$ relation for these galaxies in Figure 5a, and for reference, the fit to the relation for early-type galaxies is also shown (joint fit to our sample and S07 extended sources).  Above $L_{\ast}$, the detected late-type galaxies follow the early-type $L_K-L_X$ relation very well, but below this they are significantly more X-ray luminous for their $K_s$-band luminosities (or stellar mass).  The upper limits for the undetected late-type galaxies do not appear to be significantly underluminous compared to the detections, but deeper observations could reveal whether these galaxies are significantly lacking in hot gas or star formation compared to the detected galaxies.  In addition, there are four faint, $L_{K} \sim 10^{10} L_{K,\odot}$ galaxies with significant X-ray luminosities detected in the S07 sample, while we do not detect any similar late-type galaxies in our groups.  The spectral coverage for both the cluster and group samples comes from the literature and is very varied; however, all but two of our groups have either deep spectral coverage from the work of Zabludoff \& Mulchaey (1998) or are covered by the Sloan Digital Sky Survey (SDSS), so we do not believe that we are missing many faint late-type galaxies from our sample.  On the other hand, our sample size is smaller than that of S07 and the galaxy densities in groups are significantly lower than in clusters, so we can not be certain whether clusters host a higher fraction of these sources or if we simply miss them due to smaller statistics.

The bright X-ray luminosities of these $K_s$-band faint cluster galaxies most likely stems from significant star formation in these galaxies, and in fact, similar to S07, we find that the X-ray luminosities of late-type galaxies correlate significantly better with B-band luminosity (Figure 5b) than with $K_s$-band luminosity.  Similarly, early observations with \textit{Einstein} showed that the X-ray luminosity of late-type galaxies correlates more strongly with B-band luminosity than with H-band luminosity and that the $L_B-L_X$ relation for late-type galaxies is significantly flatter than for early-type galaxies (Fabbiano \& Trinchieri 1985; Trinchieri \& Fabbiano 1985).  In Figure 5b, we also show the O'Sullivan et al. (2001) fits (fixed slope of one) to the Fabbiano, Kim \& Trinchieri (1992) \textit{Einstein} samples of local Sb (dashed) and Sc (dot-dashed) galaxies, corrected to our luminosity band.  These fits agree fairly well with the group and cluster galaxies.  Of the S07 cluster galaxies with $L_{K} \sim 10^{10} L_{K,\odot}$, one is a known starburst in A3627 with a significant X-ray tail noted by S07.  Two more of these galaxies lie in the cluster A1367 and have significant H$\alpha$ detections, also indicating ongoing star-formation (Iglesias-P{\'a}ramo et al. 2002).  As noted by S07, A1367 also hosts another X-ray bright, starburst galaxy, UGC6697, at $L_{K} \sim L_{\ast}$, whose X-ray luminosity places it significantly above the $L_K-L_X$ relation.  This galaxy also shows an X-ray tail and signs of interaction with the ICM (Sun \& Vikhlinin 2005).  Intriguingly, A3627 and A1367 are both merging clusters.  In particular, A1367 includes infalling groups of star-forming galaxies (e.g. Cortese et al. 2004; Sakai et al. 2002).  The fourth X-ray luminous, $L_{K} \sim 10^{10} L_{K,\odot}$ cluster galaxy is a radio galaxy in the Perseus cluster which is not extended in the Chandra data but is classified as a halo in S07 based on having a significantly steep power law slope.

\begin{figure}
\epsscale{0.8}
\plotone{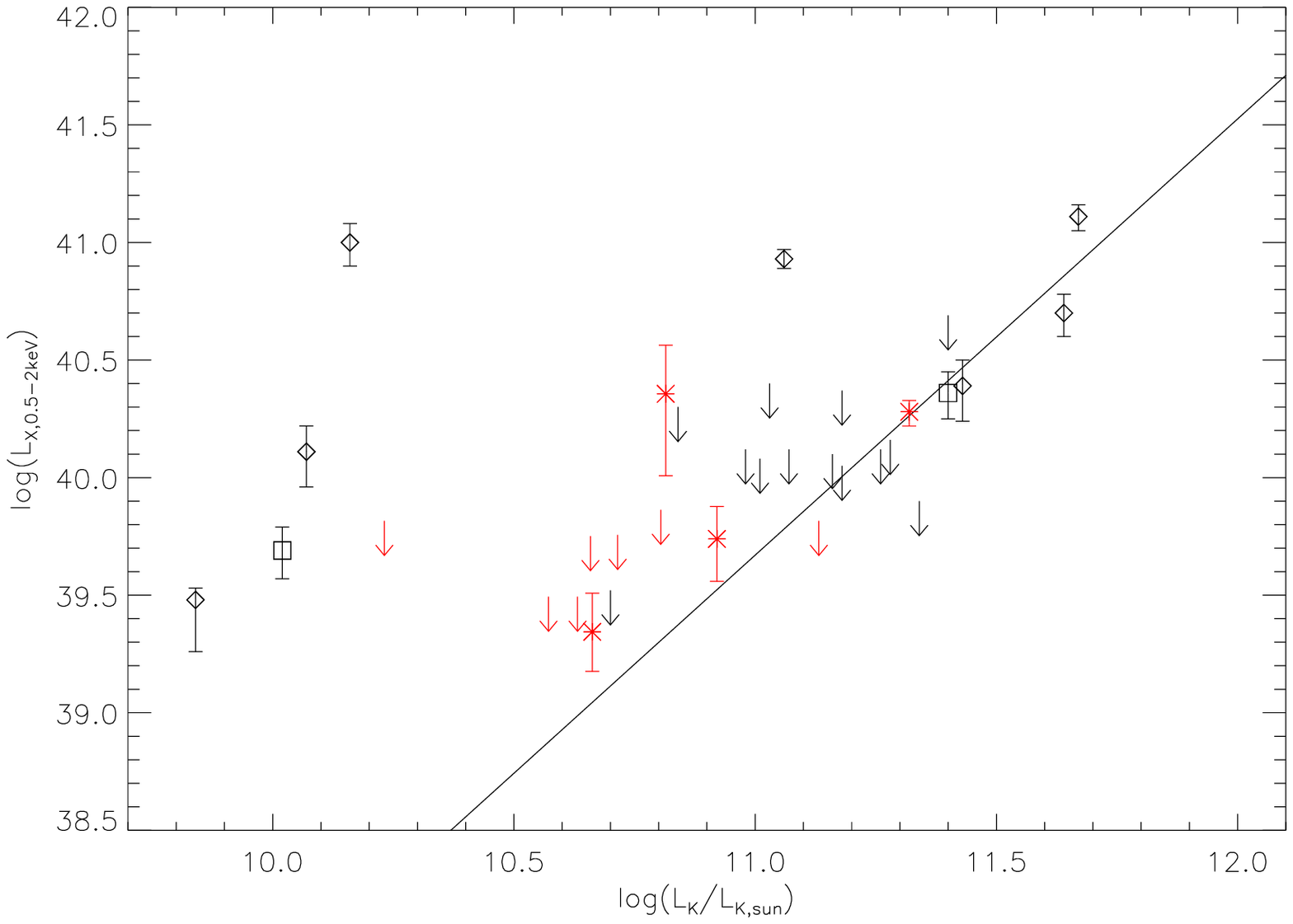}
\plotone{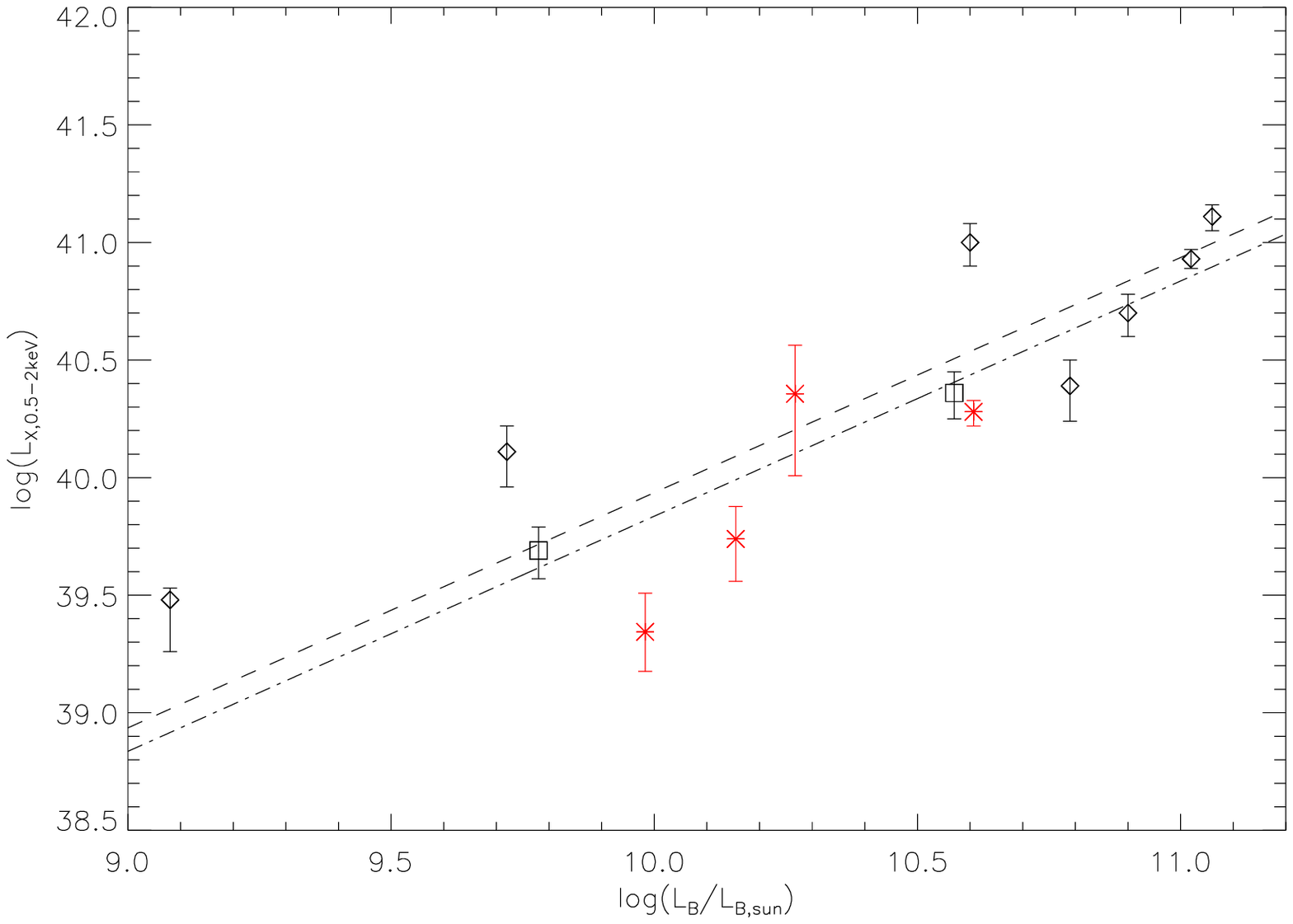}
\caption{ Top: $L_K-L_{0.5-2 keV}$ relation of late-type galaxies in groups (red) and clusters (black).  For the S07 cluster sample, diamonds indicate extended X-ray halos and squares are halos identified by their spectra.  For comparison, the solid line shows the best-fit relation for early-type galaxies (group plus cluster) detected as extended. Bottom: $L_B-L_{0.5-2 keV}$ relation of late-type galaxies in groups and clusters. Lines show the O'Sullivan et al. (2001) fits to the Fabbiano, Kim \& Trinchieri (1992) \textit{Einstein} samples of local Sb (dashed) and Sc (dot-dashed) galaxies. }
\end{figure}

\subsection{ Tails and Mergers }

Among our 21 group galaxies detected with extended X-ray emission, we find that four show indications of tails or asymmetries in their X-ray distribution oriented away from the group (NGC380, NGC508, ARK066, and NGC6265).  All of these are early-type galaxies.  The first three feature only small, perhaps insignificant tails extending $\sim$6 kpc, but NGC6265 has a long X-ray tail extending 40-50 kpc, a strong indication of gas stripping.  NGC6265 is an S0 galaxy lying 250 kpc from the X-ray center of the luminous group NGC6269.  Figure 6 shows an overlay of the Chandra contours on the SDSS image of this galaxy.  When fitting the X-ray spectrum of this source, we found that we could not constrain well the photon index of the possible power law component, but adding this component did improve the fit, so we fixed the photon index at 1.7.  NGC6265 then has an X-ray temperature of $0.56^{+0.05}_{-0.06}$ keV and a metallicity of $0.6 \pm 0.2$.  S07 find long X-ray tails associated with one early-type galaxy and two late-type, starburst galaxies (Sun \& Vikhlinin 2005; Sun et al. 2006).  NGC6265 and two of the three galaxies in S07 with long tails have $L_K \sim L_{\ast}$ or above, indicating that fairly massive galaxies can be stripped.  In total, a small fraction of cluster and group galaxies show long tails, $\sim$ 5\% of galaxies detected with extended halos and $\sim$ 2\% of all $\geq L_{\ast}$ galaxies.  The detection of tails indicates that at least some galaxies experience significant ram-pressure/viscous stripping in dense environments, but this phase is either rare or short lived.

\begin{figure}
\epsscale{0.8}
\plotone{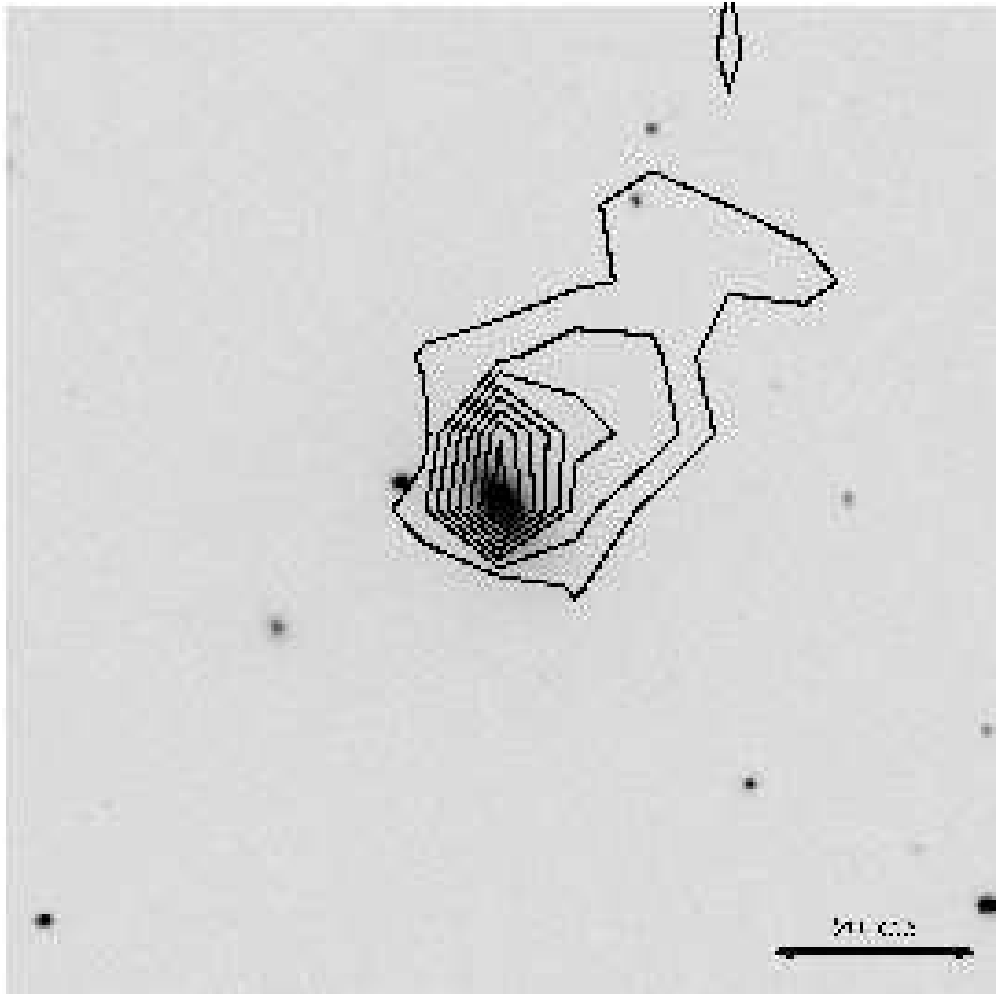}
\caption{ SDSS r band image of NGC6265 with the Chandra 0.5-2 keV contours overlaid.  This galaxy has a large X-ray tail extending 40-50 kpc. The X-ray luminous group NGC6269 lies $\sim 250$ kpc to the east.  The image is 100 kpc on a side.  The X-ray contours are derived from unsmoothed data and are linearly spaced from roughly the X-ray peak to just above the background level. }
\end{figure}

In addition, our group galaxy sample includes a few galaxy-galaxy mergers or close galaxy pairs.  For two such systems, we find indications of interaction in X-rays in the form of bridges/tails of X-ray emission between galaxies or common X-ray halos.  The first system is a merger between two active, spiral galaxies in HCG16, shown in Figure 7a.  A common X-ray halo surrounds the two galaxies and an eastern extension, which includes an X-ray point source, traces a tidal tail seen in the optical image.  The second merging system is a system of three galaxies in HCG90, two ellipticals and one spiral, shown in Figure 7b.  Optical observations reveal that the southern two galaxies are strongly interacting and more weakly interacting with the northern elliptical galaxy.  The Chandra observation reveals a common X-ray halo surrounding the southern galaxy pair, a western extension following the tail of the disrupted spiral galaxy, and a bridge of X-ray emission connecting to the northern elliptical.  These features in HCG90 were also noted by White et al. (2003) who find a large diffuse intracluster light component in the optical imaging of this group.  Tidal interactions and mergers between galaxies can morphologically transform them, tidally disrupt their outer gas and stars, and drive starbursts/AGN activity which blow out large amounts of material.  Sansom et al. (2006), in fact, find that young early-type galaxies/post mergers have low X-ray luminosities for their $K_s$-band luminosities.  The two examples presented here lie at the centers of compact groups, which represent a somewhat specialized environment, but may be a common evolutionary phase in the formation of the central, brightest group and cluster galaxies.

\begin{figure}
\epsscale{0.6}
\plotone{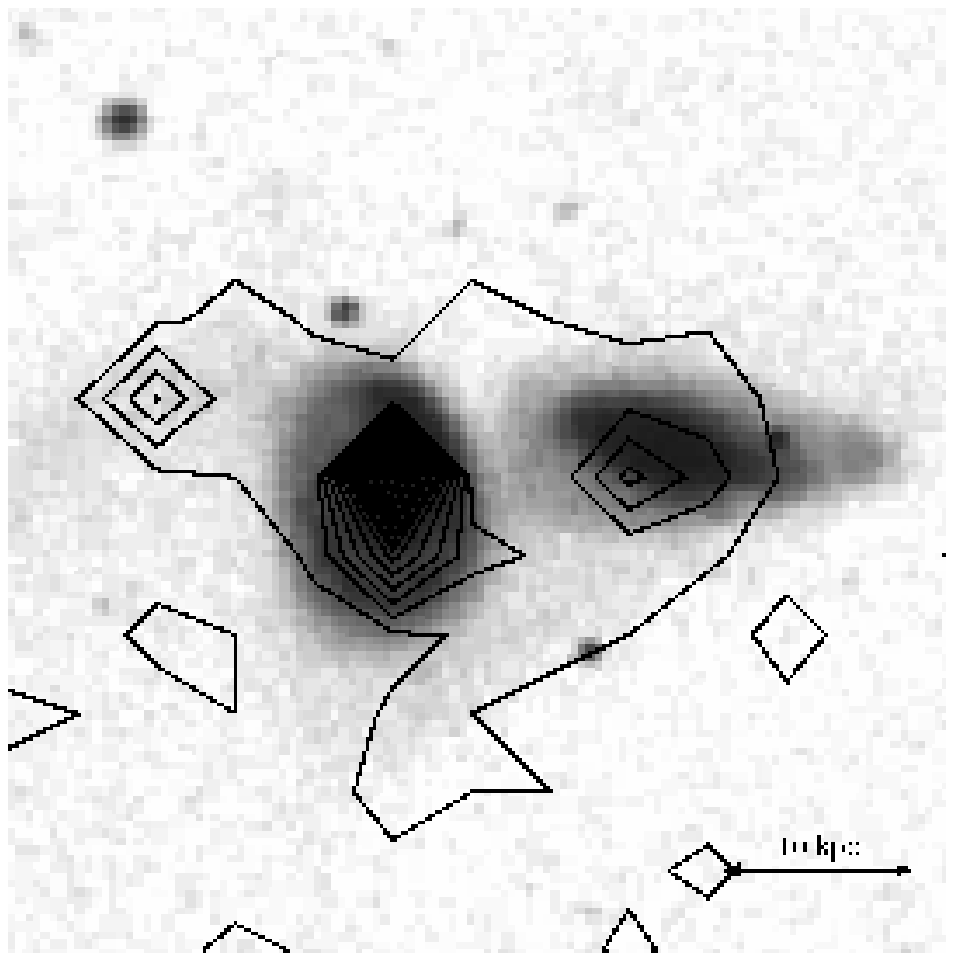}
\plotone{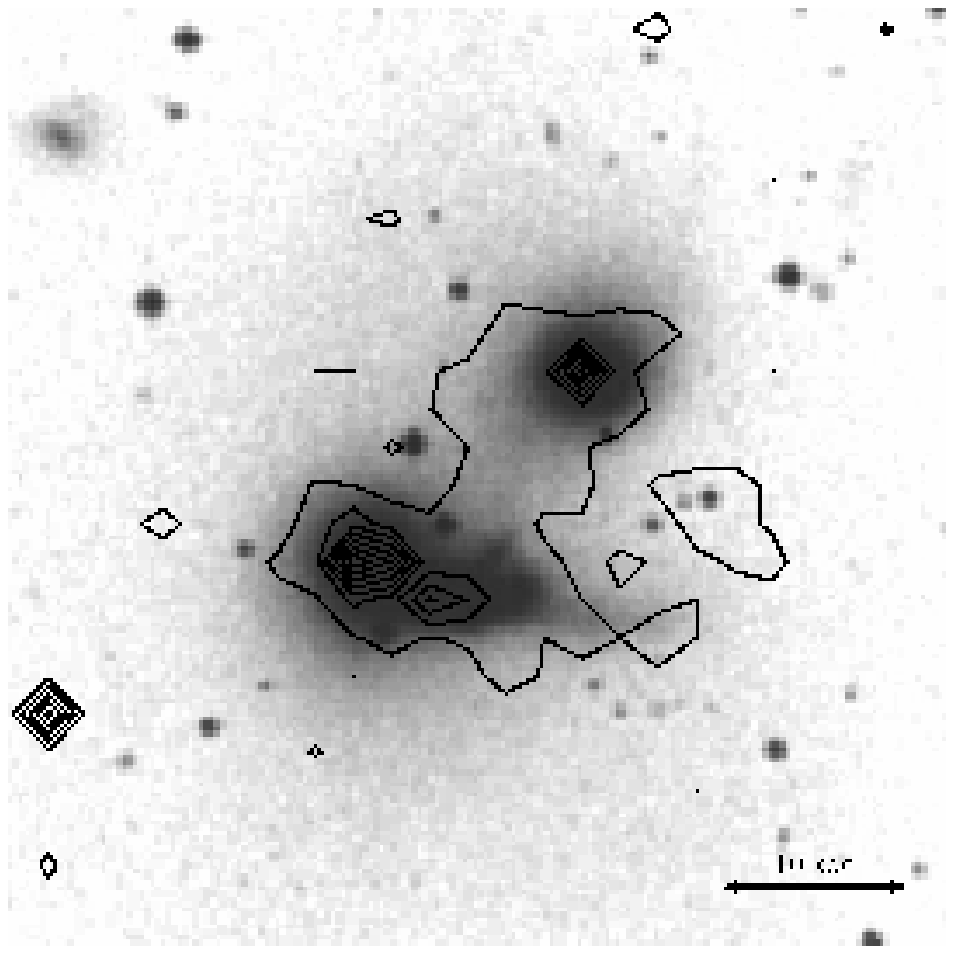}
\caption{ DSS images of galaxy-galaxy mergers in the group sample showing signs of interaction in X-rays overlaid with the Chandra 0.5-2 keV contours. Top: Merger of two spirals in HCG16.  Bottom: Merger of three galaxies in HCG90. The images are 50 kpc on a side.  The X-ray contours are derived from unsmoothed data and are linearly spaced from roughly the X-ray peak to just above the background level. }
\end{figure}

In addition to the galaxies discussed here, X-ray observations of $\sim$ 10 other cluster or group galaxies have been found to show tails or wakes indicative of ram-pressure stripping (e.g. Forman et al. 1979; Rangarajan et al. 1995; Irwin \& Sarazin 1996; Biller et al. 2004; Trinchieri et al. 1997; Kim et al. 2007; Sakelliou et al. 2005; Wang et al. 2004; Scharf et al. 2005; Machacek et al. 2005a, 2006; Rasmussen et al. 2006) and/or galaxy-galaxy interactions (Machacek et al. 2005b, 2007), making these systems rare but not uncommon.

\section{ SUMMARY AND DISCUSSION }

Using archival \textit{Chandra} observations of 13 groups, we search for hot gas halos surrounding group member galaxies.  Here we only consider satellite galaxies or BGGs which do not lie at the center of a luminous group-size X-ray halo.  We find that a large fraction of bright, early-type galaxies in groups have extended X-ray emission, indicating that they retain significant hot gas halos even in these dense environments.  In particular, we detect hot gas halos in $\sim$80\% of $L_K > L_{\ast}$ early-type galaxies.

We compare the X-ray halos of group galaxies to cluster galaxies, a number of which are also found to host significant hot gas halos (S07).  Within the errors, we do not find a significant difference in the $L_K-L_X$ relation for detected group and cluster early-type galaxies, indicating that there are not large differences in the size of the hot halos for those galaxies which host significant halos in these environments.  However, we detect a significantly higher fraction of $L_K > L_{\ast}$ early-type galaxies in groups compared to clusters, indicating that a larger fraction of galaxies in clusters experience significant stripping of their hot gas.  Ram-pressure stripping is, in fact, expected to be more effective in clusters versus groups where the ICM density is higher and the galaxy velocities are larger, while the detected galaxies are likely those who, due to their orbits and accretion times, have experienced only mild ram pressure.  We also find that group galaxies tend to have lower halo temperatures than cluster galaxies at the same near-IR luminosity or stellar velocity dispersion indicating typically less heating of the gas relative to the stars, particularly for the higher velocity dispersion galaxies in groups.

Our sample includes significantly fewer late-type galaxies, but we do detect 4 out of 11.  Combined with the late-type galaxies in the cluster sample of S07, the X-ray emission of late-type galaxies does not correlate as strongly with $K_s$-band luminosity (S07) as it does for early-type galaxies.  Brighter late-types ($L_K > L_{\ast}$) appear to follow the $L_K-L_X$ relation for early-type galaxies, while the fainter late-type galaxies are significantly more X-ray luminous than this relation.  In particular, clusters, at least, contain a population of $K_s$-band faint, X-ray luminous, star-forming galaxies.  The X-ray luminosity of group and cluster late-type galaxies correlates much more strongly with B-band luminosity, supporting the theroy that the X-ray emission is related to star formation.  Intriguingly, the detected group and cluster late-type galaxies do not appear to have singificantly lower $L_X/L_B$ than the overall local spiral population as observed by \textit{Einstein}.  The upper limits for the undetected late-types do not appear to be significantly underluminous compared to the detected galaxies, and deeper observations are needed to establish if these galaxies are X-ray deficient.

The fact that many galaxies in groups and clusters retain significant hot gas halos is in qualitative agreement with recent hydrodynamic simulations which show that for typical galaxy structures and orbits, galaxies falling in to groups and clusters maintain $\sim$30\% of their hot halo gas after 10 Gyr (McCarthy et al. 2007), with the amount of stripping depending signficantly on the orbital history.  Our result that a significantly higher fraction of galaxies are detected in groups versus clusters is also in qualitative agreement with the predictions of recent semi-analytical simulations based on the Millenium Simulations, which find that a higher fraction of galaxies in massive clusters versus poor clusters experience strong ram pressure at some point in their history (Br\"{u}ggen \& De Lucia 2007).  However, these simulations also predict that nearly all galaxies in cluster and group cores have experienced strong ram pressure, seemingly at odds with the large number of detected X-ray halos.

These models also predict that there should be a strong radial trend in the stripping experienced by group and cluster galaxies, but we find no significant trend of the $L_K-L_X$ relation or detections versus non-detections with radius.  Here we face limitations of both the sample size and the radial coverage of the \textit{Chandra} data, which may limit our ability to detect such a trend.  Our detected group galaxies are limited in radius to $r \leq 0.3 r_{500}$.  Comparing to field galaxies in the \textit{ROSAT} observed sample of Ellis \& O'Sullivan (2006), we do find that field early-type galaxies have a significantly higher X-ray luminosity at fixed $L_K$ than group and cluster galaxies.  We find a significant offset in the $L_K-L_X$ relation between group and field galaxies even when we assume that all of the X-ray emission from detected sources originates from a thermal component, as was done by Ellis \& O'Sullivan (2006).  This result may confirm theoretical expectations that the hot halos of galaxies in dense environments are significantly stripped compared to field galaxies, but analysis of a field sample with \textit{Chandra} is needed to confirm that the \textit{ROSAT} luminosities are not significantly enhanced by point source or background contamination.

It is, however, clear that ram pressure and tidal stripping of hot galactic gas does occur in at least some galaxies in dense environments.  We find that $\sim$ 5\% of galaxies detected with extended halos and $\sim$ 2\% of all $\geq L_{\ast}$ galaxies in groups and clusters show long X-ray tails indicative of ram-pressure/viscous stripping.  We additionally find two systems of merging galaxies in our groups that show indications of tidal interaction in their X-ray gas.  These cases add to the $\sim$ 10 other cluster/group galaxies in the literature found to show X-ray signatures of ram pressure and/or tidal stripping (Forman et al. 1979; Rangarajan et al. 1995; Irwin \& Sarazin 1996; Biller et al. 2004; Trinchieri et al. 1997; Kim et al. 2007; Sakelliou et al. 2005; Wang et al. 2004; Scharf et al. 2005; Machacek et al. 2005a, 2005a, 2006, 2007; Rasmussen et al. 2006).  These systems include both early and late-type galaxies.  We also find galaxies in both groups and clusters which are significantly underluminous (both detections and non-detections) compared to the $L_K-L_X$ relation, with a higher fraction of these galaxies in clusters.

A picture is therefore emerging that stripping of hot galactic gas through both ram pressure and tidal forces does occur in groups and clusters, but the frequency or efficiency of such events must be moderate enough to allow the hot gas halos in a large fraction of bright galaxies to survive even in group and cluster cores.  Our results also indicate that a larger fraction of cluster galaxies have experienced significant gas stripping compared to galaxies in groups.  The question remains whether these processes contribute significantly to galaxy evolution or ICM enrichment.  This question can be addressed with future simulations using these observations as a baseline for the hot gas content of galaxies in groups and clusters.

\acknowledgments
We would like to thank the referee for their insightful comments on our paper.  T.E.J. is grateful for support from the Alexander F. Morrison Fellowship, administered through the University of California Observatories and the Regents of the University of California.  J.S.M. acknowledges partial support for this work from NASA grant NNG04GC846.

\begin{deluxetable}{lcccccc}
\tablecaption{ Group Sample }
\tablewidth{0pt}
\tablecolumns{6}
\tablehead{
\colhead{Group} & \colhead{Obs. ID} & \colhead{Instrument} &\colhead{Exposure\tablenotemark{a}} & \colhead{Redshift\tablenotemark{b}} & \colhead{$\sigma_v$\tablenotemark{c}} & \colhead{$L_X$\tablenotemark{b}}\\
\colhead{} & \colhead{} & \colhead{} & \colhead{(secs)} & \colhead{} & \colhead{(km s$^{-1}$)} & \colhead{($10^{42}$ ergs s$^{-1}$)}
}
\startdata
NGC 383 &2147 &ACIS-S &41802 &0.0173 &450 &5.25 \\
NGC 507 &2822 &ACIS-I &43539 &0.0170 &608 &8.91 \\
NGC 533 &2880 &ACIS-S &30959 &0.0181 &439 &2.34 \\
NGC 741 &2223 &ACIS-S &28299 &0.0185 &453 &1.38 \\
HCG 16 &923 &ACIS-S &12347 &0.0131 &80 &0.0550 \\
HCG 42 &3215 &ACIS-S &31042 &0.0128 &282 &0.389 \\
NGC 3557 &3217 &ACIS-I &33773 &0.0095 &300 &0.132 \\
NGC 4073 &3234 &ACIS-S &27915 &0.0201 &565 &11.0 \\
NGC 4325 &3232 &ACIS-S &23756 &0.0254 &376 &6.31 \\
HCG 62 &921 &ACIS-S &47922 &0.0145 &418 &4.90 \\
NGC 5171 &3216 &ACIS-S &33650 &0.0232 &494 &2.24 \\
NGC 6269 &4972 &ACIS-I &39525 &0.0353 &574 &15.8 \\
HCG 90 &905 &ACIS-I &49528 &0.0085 &131 &0.0398 \\
\enddata
\tablenotetext{a}{Net exposure time after flare filtering.}
\tablenotetext{b}{Group redshift and X-ray luminosity are taken from Mulchaey et al. (2003).}
\tablenotetext{c}{Group velocity dispersion is taken from Osmond \& Ponman (2004) except for NGC507 and NGC6269, which are calculated from the NED galaxy catalogs.}
\end{deluxetable}

\begin{deluxetable}{lcccccccccc}
\tablecaption{ X-ray Detected Group Members }
\tablewidth{0pt}
\rotate
\tabletypesize{\footnotesize}
\tablecolumns{11}
\tablehead{
\colhead{Group} & \colhead{Galaxy} & \colhead{z\tablenotemark{a}} & \colhead{log($L_K/L_{K,\odot}$)} & \colhead{$L_{0.5-2keV}$\tablenotemark{b}} & \colhead{$L_{X,bol}$\tablenotemark{b}} & \colhead{kT} & \colhead{$\Gamma$} & \colhead{Morph.\tablenotemark{c}} & \colhead{R\tablenotemark{d}} & \colhead{notes\tablenotemark{e}} \\
\colhead{} & \colhead{} & \colhead{} & \colhead{} & \colhead{($10^{39}$ egs s$^{-1}$)} & \colhead{($10^{39}$ egs s$^{-1}$)} & \colhead{(keV)} & \colhead{} & \colhead{} & \colhead{(kpc)} & \colhead{} 
}
\startdata
NGC 383 &NGC 0382 &0.01744 &11.13 &$4.74^{+1.54}_{-1.29}$ &7.23 &$0.45^{+0.09}_{-0.09}$ &0.0 &E &11.82 &E \\
NGC 383 &NGC 0380 &0.01476 &11.27 &$44.28^{+2.22}_{-2.03}$ &64.58 &$0.63^{+0.02}_{-0.02}$ &$1.71^{+0.22}_{-0.18}$ &E2 &81.21 &E \\
NGC 383 &NGC 0385 &0.01659 &11.11 &$9.05^{+2.46}_{-2.25}$ &14.00 &$0.48^{+0.09}_{-0.12}$ &1.7 &SA0 &113.17 &E \\
NGC 383 &NGC 0379 &0.01861 &11.42 &$17.85^{+2.95}_{-2.67}$ &25.92 &$0.53^{+0.07}_{-0.07}$ &$2.10^{+0.39}_{-0.17}$ &S0 &153.08 &E \\
NGC 383 &NGC 0384 &0.01412 &10.93 &$10.26^{+7.28}_{-6.96}$ &207.24 &$0.22^{+0.03}_{-0.03}$ &1.7 &E3 &124.28 &E \\
NGC 507 &NGC 0508 &0.01843 &11.28 &$48.44^{+25.09}_{-28.87}$ &1664.50 &$0.19^{+0.02}_{-0.02}$ &1.7 &E0 &32.97 &E \\
NGC 507 &NGC 0504 &0.01410 &10.91 &$<$2.69 &$<$4.06 &0.7 &0.0 &S0 &68.69 &E \\
NGC 507 &IC1687 &0.01674 &10.76 &$<$5.49 &$<$8.30 &0.7 &0.0 &E3 &91.98 &E \\
NGC 507 &ARK039 &0.01675 &10.63 &$<$6.56 &$<$9.91 &0.7 &0.0 &S &106.22 &E \\
NGC 533 &2MJ0125+0145 &0.01716 &10.11 &$<$4.46 &$<$6.74 &0.7 &0.0 &E &87.52 &P \\
NGC 741 &NGC 0742 &0.01991 &10.82 &$<$5.85 &$<$8.77 &0.7 &$2.14^{+0.07}_{-0.07}$ &cE0 &19.46 &P \\
NGC 741 &ARK066 &0.02113 &10.85 &$6.17^{+1.21}_{-1.21}$ &9.32 &0.7 &0.0 &E &37.66 &E \\
NGC 741 &ARK065 &0.01767 &10.64 &$<$1.86 &$<$2.81 &0.7 &0.0 &E &71.00 &U \\
HCG 16 &NGC 0838 &0.01284 &10.95 &$22.58^{+3.28}_{-3.35}$ &33.13 &$0.64^{+0.05}_{-0.05}$ &$2.20^{+0.14}_{-0.10}$ &SA, pec &0.00 &E, Sbrst \\
HCG 16 &NGC 0835 &0.01359 &11.32 &$19.05^{+2.21}_{-2.49}$ &27.67 &$0.54^{+0.05}_{-0.06}$ &$0.37^{+0.20}_{-0.13}$ &SAB, pec &57.75 &E, LINR \\
HCG 16 &NGC 0833 &0.01289 &11.06 &$<$5.24 &$<$7.96 &0.7 &1.7 &Sa, pec &69.92 &P, Sy2 \\
HCG 16 &NGC 0839 &0.01292 &10.92 &$5.49^{+2.05}_{-1.87}$ &9.48 &$0.86^{+0.12}_{-0.08}$ &1.7 &S, pec &39.83 &E, Sy2 \\
HCG 42 &NGC 3096 &0.01410 &10.71 &$<$1.27 &$<$2.00 &0.7 &0.0 &SB0 &81.23 &U \\
HCG 42 &MCG-03-26-006 &0.01336 &10.76 &$<$0.67 &$<$1.03 &0.7 &0.0 &E &20.93 &U \\
NGC 3557 &NGC 3557 &0.01019 &11.18 &$41.31^{+37.76}_{-14.48}$ &171.71 &$0.25^{+0.05}_{-0.03}$ &$1.97^{+0.20}_{-0.27}$ &E3 &0.00 &E \\
NGC 3557 &NGC 3564 &0.00940 &11.01 &$3.51^{+1.24}_{-1.11}$ &5.55 &$0.52^{+0.15}_{-0.18}$ &1.7 &S0 &89.10 &E \\
NGC 3557 &NGC 3568 &0.00815 &10.81 &$22.72^{+13.85}_{-12.53}$ &445.94 &$0.20^{+0.03}_{-0.02}$ &$1.26^{+0.30}_{-0.26}$ &SB &115.48 &E \\
NGC 4073 &NGC 4063 &0.01640 &10.82 &$<$4.08 &$<$6.17 &0.7 &0.0 &S0 &121.23 &P \\
NGC 4325 &NGC 4320 &0.02668 &11.13 &$<$11.30 &$<$17.02 &0.7 &0.0 &S &155.21 &P \\
HCG 62 &NGC 4761 &0.01478 &10.73 &$<$1.20 &$<$1.80 &0.7 &0.0 &E &20.08 &E \\
HCG 62 &NGC 4759NED01 &0.01188 &10.85 &$1.63^{+0.55}_{-0.53}$ &2.88 &$0.32^{+0.08}_{-0.05}$ &1.7 &S0, pec &5.68 &E \\
NGC 5171 &NGC 5171 &0.02294 &11.35 &$13.05^{+2.24}_{-2.53}$ &21.54 &$0.83^{+0.15}_{-0.11}$ &1.7 &S0 &49.68 &E \\
NGC 5171 &SDSSJ1329+1144 &0.02335 &10.50 &$<$3.73 &$<$5.76 &0.7 &0.0 &(E0) &46.55 &U \\
NGC 5171 &NGC 5179 &0.02413 &11.18 &$8.35^{+2.11}_{-1.92}$ &13.94 &$0.40^{+0.12}_{-0.09}$ &1.7 &S0 &52.36 &E \\
NGC 5171 &NGC 5176 &0.02368 &11.12 &$3.20^{+1.69}_{-1.55}$ &4.71 &0.7 &1.7 &S0 &33.33 &E \\
NGC 5171 &NGC 5177 &0.02157 &10.77 &$<$2.21 &$<$3.34 &0.7 &0.0 &S0 &55.08 &U \\
NGC 6269 &2MJ1657+2755 &0.03459 &10.23 &$<$10.01 &$<$15.09 &0.7 &0.0 &(Sa/b) &177.94 &U \\
NGC 6269 &NGC 6265 &0.03245 &11.37 &$111.32^{+15.67}_{-11.65}$ &163.94 &$0.56^{+0.05}_{-0.06}$ &1.7 &S0 &250.50 &E \\
HCG 90 &NGC 7173 &0.00833 &10.89 &$0.94^{+0.26}_{-0.23}$ &1.52 &$0.84^{+0.15}_{-0.11}$ &$-0.02^{+0.23}_{-0.33}$ &E, pec &0.01 &E \\
HCG 90 &NGC 7176 &0.00838 &11.27 &$1.94^{+0.59}_{-0.61}$ &3.39 &$0.37^{+0.07}_{-0.08}$ &$1.69^{+0.15}_{-0.12}$ &E, pec &15.26 &E \\
HCG 90 &NGC 7174 &0.00887 &10.66 &$2.21^{+1.02}_{-0.71}$ &3.50 &$0.54^{+0.11}_{-0.16}$ &1.7 &Sab, pec &14.66 &E \\
HCG 90 &NGC 7172 &0.00868 &11.18 &$<$1.41 &$<$2.11 &0.7 &1.2 &Sa, pec &66.77 &E, Sy2 \\
\enddata
\tablenotetext{a}{Galaxy redshift from NED.}
\tablenotetext{b}{0.5-2 keV and bolometric luminosities of the thermal component from the spectral fit.}
\tablenotetext{c}{Galaxy morphologies are from NED unless in parentheses in which case the morphology was visually determined from SDSS r band images.}
\tablenotetext{d}{Distance from the ICM peak or the distance from the central galaxy in X-ray faint groups (HCG16, NGC3557, and HCG90).}
\tablenotetext{e}{(E) Extended X-ray source, (P) Point source, (U) Fewer than 40 counts, unknown extent.  Also noted are starburst, Seyfert, and LINER galaxies from NED.}
\end{deluxetable}

\begin{deluxetable}{lccccccc}
\tablecaption{ Undetected Group Members }
\tablewidth{0pt}
\rotate
\tablecolumns{8}
\tabletypesize{\small}
\tablehead{
\colhead{Group} & \colhead{Galaxy} & \colhead{z\tablenotemark{a}} & \colhead{log($L_K/L_{K,\odot}$)} & \colhead{$L_{0.5-2keV}$} & \colhead{$L_{X,bol}$} & \colhead{Morph.\tablenotemark{b}} & \colhead{R\tablenotemark{c}} \\
\colhead{} & \colhead{} & \colhead{} & \colhead{} & \colhead{($10^{39}$ egs s$^{-1}$)} & \colhead{($10^{39}$ egs s$^{-1}$)} & \colhead{} & \colhead{(kpc)}
}
\startdata
NGC 383 &NGC 0386 &0.01853 &10.68 &$<$3.76 &$<$5.64 &E3 &74.90 \\
NGC 383 &NGC 0375 &0.01953 &10.63 &$<$1.88 &$<$2.82 &E2 &132.40 \\
NGC 383 &NGC 0373 &0.01835 &10.54 &$<$1.75 &$<$2.63 &E &188.20 \\
NGC 507 &NGC 0503 &0.01975 &10.81 &$<$2.73 &$<$4.09 &E &123.51 \\
NGC 507 &IC1690 &0.01546 &10.61 &$<$1.41 &$<$2.11 &S0 &119.28 \\
NGC 4073 &CGCG013-058 &0.02386 &10.83 &$<$2.82 &$<$4.22 &SA0 &101.50 \\
NGC 4073 &NGC 4139 &0.01868 &10.93 &$<$2.92 &$<$4.37 &SB0 &134.92 \\
NGC 4073 &NGC 4075 &0.02184 &11.30 &$<$2.84 &$<$4.26 &SA0 &289.27 \\
NGC 4325 &VPC0243 &0.02608 &10.55 &$<$3.82 &$<$5.71 &E &277.76 \\
NGC 5171 &CGCG072-088 &0.02472 &10.71 &$<$5.71 &$<$8.54 &(Sb) &187.42 \\
NGC 5171 &2MJ1329+1148 &0.02640 &10.57 &$<$3.11 &$<$4.66 &(Sc) &134.95 \\
NGC 6269 &2MJ1657+2752 &0.03559 &10.56 &$<$7.54 &$<$11.26 &(Epec) &54.19 \\
NGC 6269 &2MJ1657+2756 &0.03839 &10.73 &$<$13.99 &$<$20.86 &(E/S0) &235.13 \\
NGC 6269 &2MJ1658+2748 &0.03400 &10.50 &$<$5.10 &$<$7.62 &(E) &213.90 \\
NGC 6269 &2MJ1657+2748 &0.03699 &10.70 &$<$5.67 &$<$8.46 &(E) &260.81 \\
NGC 6269 &2MJ1657+2755 &0.03748 &10.50 &$<$6.46 &$<$9.64 &(E) &317.15 \\
NGC 6269 &2MJ1658+2757 &0.03328 &10.80 &$<$5.15 &$<$7.70 &(Sa) &313.83 \\
NGC 6269 &GIN631 &0.03653 &10.80 &$<$7.29 &$<$10.87 &S &406.68 \\
NGC 6269 &2MJ1658+2753 &0.03577 &10.66 &$<$5.64 &$<$8.42 &(SB) &409.82 \\
NGC 6269 &NGC 6270 &0.03242 &11.13 &$<$6.54 &$<$9.79 &E/S0 &394.39 \\
\enddata
\tablenotetext{a}{Galaxy redshift from NED.}
\tablenotetext{b}{Galaxy morphologies are from NED unless in parentheses in which case the morphology was visually determined from SDSS r band images.}
\tablenotetext{c}{Distance from the ICM peak or the distance from the central galaxy in X-ray faint groups (HCG16, NGC3557, and HCG90).}
\end{deluxetable}

\begin{deluxetable}{lcccc}
\tablecaption{ $L_X - L_K$ Relation }
\tablewidth{0pt}
\tablecolumns{5}
\tablehead{
\colhead{} & \colhead{Detected} & \colhead{} & \colhead{With Non-Detections} & \colhead{} \\
\colhead{Sample} & \colhead{A} & \colhead{B} & \colhead{A} & \colhead{B}
}
\startdata
Group Galaxies &$39.63 \pm 0.14$ &$2.74 \pm 0.63$ &$39.33$ &$2.90 \pm 0.71$ \\
Cluster Galaxies &$39.73 \pm 0.13$ &$1.57 \pm 0.28$ &- &-\\
(S07 extended) & & & & \\
Cluster Galaxies &$39.75 \pm 0.05$ &$1.49 \pm 0.14$ &$39.32$ &$1.64 \pm 0.14$ \\
(S07, all) & & & & \\
Group + Cluster &$39.67 \pm 0.10$ &$1.86 \pm 0.23$ &- &-\\
(S07 extended) & & & & \\
Field Galaxies &$40.34 \pm 0.18$ &$1.85 \pm 0.44$ &- &-\\
Group Galaxies &$39.76 \pm 0.09$ &$2.16 \pm 0.32$ &- &-\\
(assumed thermal) & & & & \\
\enddata
\end{deluxetable}

\end{document}